\documentclass[twocolumn,prl,aps]{revtex4}
\usepackage{graphicx}
\renewcommand{\vec}[1]{{\mathbf{#1}}}
\newcommand{\beq}{\begin{eqnarray}}
\newcommand{\eeq}{\end{eqnarray}}
\begin{document}

\title{Squaring the Triangle:  Insulating Ground State of $Na_{0.5}CoO_{2}$}
\author{Ting-Pong Choy, Dimitrios Galanakis, Philip Phillips}
\affiliation{Loomis Laboratory of Physics, University of Illinois at
Urbana-Champaign, 1110 W. Green St., Urbana, IL, 61801}

\begin{abstract}
We demonstrate that at a filling of $n=1.5$, an hexatic insulating
state obtains in the extended Hubbard model on a triangular lattice.
Composed of two tetragonal sublattices with fillings of $n=1$ and
$n=2$, the insulating state is charge ordered and possesses an
antiferromagnetic superlattice with dimension $a\times\sqrt{3}$. Two
distinct energy scales arise in our model, a charge gap for the
insulator and the effective exchange interaction in the
antiferromagnet. Our model is capable of explaining both qualitatively and quantitatively the Hall coefficient including the sign change, the temperature dependence of the resistivity and the
persistence of antiferromagnetism above the insulating state.
\end{abstract}

\maketitle The cobaltates\cite{Takada}, $Na_{x}CoO_{2}$ are layered
anisotropic materials that share many of the complexities of the
high-temperature copper-oxide superconductors.  They consist of
insulating layers of $Na^{+}$ ions separated by $CoO_{2}$ planes.
As in the cuprates, superconductivity in the hydrated
cobaltates\cite{Takada}, $Na_xCoO_2\cdot yH_2 O$,arises from doping
a half-filled Mott insulating state. In the cobaltates, the
half-filled band arises in the parent material, $Na_0CoO_2$, from a
complete filling of each of the $e_{g}(t_{2g})$ orbitals while only
a single electron resides in the $d_{z^{2}}$
state\cite{baskaran,palee}. However, the underlying lattice is triangular as opposed to the square lattice of the cuprates.  The concentration, $x$, measures the
number of Co$^{+4}$ ions converted to Co$^{+3}$ ions for every Na
atom added. Superconductivity\cite{Takada} in
 $Na_{x}CoO_{2}\cdot y H_{2}O$, is confined to
a dome-like region as a function of 
the effective cobalt valence, which, in this case,
is determined\cite{arg} by the net
 hydronium and sodium ion contents

While the cobaltates are generally metallic, they exhibit a novel insulating
state\cite{Princeton} at $x=0.5$. At present,
the insulating behaviour at $x=0.5$ remains a mystery.  We propose that the key
mechanism driving the insulating state is the elimination of
magnetic frustration by the formation of a charge-ordered state. Two
distinct energy scales exist in the charge-ordered state: 1) the
effective in-plane antiferromagnetic exchange coupling, $J$, and 2)
the energy gap in the insulator, $\Delta$.  We show that for
reasonable system parameters, $\Delta<J$ and hence the insulating
state obtains at a lower temperature than does the antiferromagnetic
(AF) transition.  Our model predicts electron-like transport below
the AF transition but hole-like transport above, as is seen experimentally\cite{Princeton}.

There are now distinctly different probes of the insulating state at
$n=1.5$. Transport measurements\cite{Princeton} find that the
resistivity at $n=1.5$ is temperature independent from $60<T<300K$
but undergoes a sharp upturn as the temperature decreases below
$53K$. Further, the Hall coefficient\cite{Princeton} changes sign
from positive to negative as the temperature is lowered through 88K.
Unlike the square lattice which is particle-hole symmetric at $n=1$,
such is not the case for the triangular lattice. Hence, the
vanishing of the Hall coefficient in the Na$_x$CoO$_2$ requires a
non-trivial electronic state.  The magnetic susceptibility $\chi$
exhibits a slight kink at $53K$, implying a phase transition.  In
addition, neutron diffraction\cite{NeutronDiff,zand} and
NMR\cite{cob2} experiments demonstrate that the Na atoms form an
orthorhombic superlattice with lattice vectors $(a\sqrt{3}{\bf
x},2a{\bf y})$. Sodium atom ordering has been promoted as the a
central ingredient in driving the insulating state.  Consistent with
such ordering is an enhancement of the thermal
conductivity\cite{Princeton} below $53K$ only at $n=1.5$.
 However, recent
experiments on hydrated $Na_{0.41}CoO_2$ in which the conductivity
has also been observed to diverge below $53K$ call this view into
question. Electrochemical measurements\cite{alt} indicate that
charge transfer between the oxonium ion ($H_3O^+$) and Co results in
a formal valence of Co of 3.5 rather than the nominal value of 3.59
obtained from the sodium content.  Consequently, the authors
conclude that the key ingredient in driving the insulating state is
the presence of Co with a formal valence of 3.5 not sodium ordering.

\begin{figure}
\includegraphics[scale=0.4]{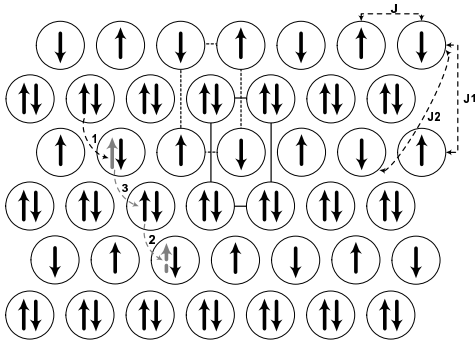}
\caption{Charge ordering of $Na_{0.5}CoO_{2}$. The circles represent
cobalt atoms. Notice that the triangular lattice is broken into two
orthorhombic lattices one with the double occupancies and one with
antiferromagnetic ordering. Moving an electron from a doubly
occupied site to a singly occupied site will increase its energy by
$J$ where $J=4t^2/U$ is the antiferromagnetic coupling constant.  A
next-nearest neighbour superexchange interaction, $J_1$, also exists
in the vertical direction. The diagonal hopping process which leaves
antiferromagnetism intact is also depicted\label{pstate}.}
\end{figure}

To understand what is special about the filling of $n=1.5$ on a
triangular lattice, consider the square lattice at $n=1$. At $n=1$,
the Coulomb repulsion energy is minimized and all magnetic
frustration is eliminated on a square lattice by forming an
antiferromagnet.  Such a state is insulating because of the charge
gap, $U$, for doubly occupying the same site.  Our key assumption
for the triangular lattice is that minimizing the Coulomb energy and
eliminating Co magnetic frustration drives the
insulating state at $n=1.5$. At a filling of $n=1.5$, all Coulomb
repulsion energy and magnetic frustration can be eliminated by
filling the lattice with alternating rows of doubly and singly
occupied sites.  Consequently, we consider the state in Fig.
(\ref{pstate}) in which the doubly and singly occupied sites occupy
the corners of orthorhombic unit cells.

Such a state can be thought of as a triangular lattice split into
two tetragonal lattices, one for the double occupancy and the other
for single occupancy with antiferromagnetic correlations. In fact,
in terms of a unit cell defined by the basis vectors, $(1,0)$ and
$(1/2,\sqrt{3}/2)$, the state proposed here preserves translational
symmetry but breaks rotational invariance by $2\pi/3$.  Hence, the
state is a hexatic nematic phase similar to that proposed for $x=1$
in the Emery model\cite{fradkin2} in which nearest neighbour Coulomb
interactions play a pivotal role. On a triangular lattice with
$n=1.5$, there is another ordered ground state, the
zig-zag state\cite{mizokawa}, which can be obtained by a zero-energy cost rearrangement of the cobalt ions in the charge-ordered state. Consequently, which state is favoured experimentally is determined by factors such as the lattice structure
and the sodium ordering pattern. Nonetheless, to establish
that either the charge-ordered or zig-zag states is an insulator
necessitates an analysis of the distinct transport channels,
which we now present.

We analyze the energetics for transport in the context of the
extended Hubbard model,
\beq\label{eq:tUV} H=-t\sum_{\langle ij\rangle} c_{i\sigma}^\dagger
c_{j\sigma}+h.c.+U\sum_i n_{i\uparrow}n_{i\downarrow}+V\sum_{\langle
ij\rangle} n_i n_j
\eeq
where $c_{i\sigma} (c_{i\sigma}^\dagger)$ annihilates (creates) an
electron with spin $\sigma$ on site i, $U$, the energy cost for
doubly occupying the same site and $V$ the nearest-neighbour Coulomb
repulsion. The importance of the nearest-neighbour interaction in
stabilizing the insulating state can be illustrated as follows.
Since along each singly occupied row, charge transport is blocked as
a result of the charge gap, $U$, it suffices to show that Fig.
(\ref{pstate}) is an insulator by demonstrating that the doubly
occupied sites are immobile. Any electron in a doubly occupied site
is surrounded by an equal number of spin up and spin down electrons
in the neighboring sites, resulting in a net zero interaction
energy. However, if an electron hops to a singly occupied site, it
will be surrounded by three double occupancies, two electrons of the
opposite spin, and one of the same spin. The net increase in the
energy will be $V+J$ where $J=4t^2/U$\cite{j3}. We will call such a
process as the creation of a doublon (on a singly-occupied row) and
holon (on a doubly-occupied row) pair which costs an energy $V+J$.
Once, a doublon-holon pair has been created, there are four distinct
transport mechanisms:

1) Holon motion along the otherwise doubly occupied superlattice:
Transport of the holon vertically along the superlattice
requires two-step hopping which does not cost any energy in the
intermediate state. The amplitude for such a process is $t$.

2) Doublon motion in the singly occupied row: It costs an energy
n$J_1$, were $n$ is the number of hops. Hence, a doublon is {\it
linearly} confined and does not propagate along a chain.

3) Transport of a doublon along a diagonal through a series of
concerted two-step hoppings as illustrated in Fig. (\ref{pstate}):
Doublon hops along the diagonal, however, do not generate any
further energy costs and hence constitute the only truly
2-dimensional transport in the antiferromagnetic phase. As doublon
motion propagates an electron, $R_H<0$. This type of diagonal
hopping process which leaves the antiferromagnet intact requires
{\bf both} the triangular lattice and the charge ordered state in
Fig. (\ref{pstate}).

4) Holon transport along either one of the diagonals. Such hopping depends on
the surrounding spin. 

\begin{figure}
\includegraphics[width=5.5cm, angle=270]{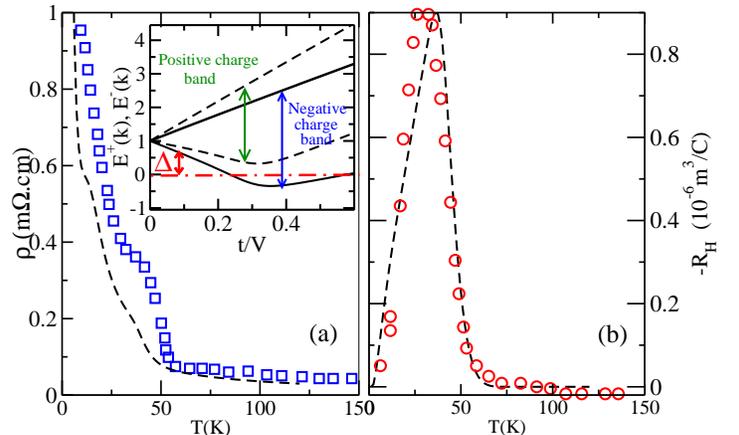}
\caption{(a) Resistivity (dashed line) and band structure  for the model in Eq.
(\ref{toy}). Two bands corresponding to negative and positive charge
quasiparticles with a gap $\Delta$ are shown. (b) Hall coefficient
for Eq. (\ref{toy}).
The experimental data  for both the resistivity (squares) and Hall coefficient (circles) are
taken from Foo, et. al.\cite{Princeton}.}\label{band}
\end{figure}

On the other hand, in the absence of antiferromagnetic
correlations, both the holon and doublon hop in their corresponding
superlattices with the same hopping matrix elements but with opposite sign.
All such transport mechanisms cease once the holon and doublon are
bound. The criterion for holon-doublon binding can be derived from a
2-dimensional model,
\beq\label{toy}
H &=& \sum\limits_{ \langle i,j \rangle } {t^d_{ij} (d_i^\dagger d_j + h.c.)} + \sum\limits_{ \langle i,j \rangle } { t^h_{ij}(h_{i\sigma}^\dagger h_{j\sigma} + h.c.)  } \\
&+& \sum_i {(V + \tilde{J}) (h_{i\sigma}^\dagger h_{i\sigma} +
d_i^\dagger d_i)} - t\sum_i {(d_i^\dagger h_{i\sigma}^\dagger +
h.c.)} \nonumber \eeq
which includes all of the distinct hopping processes delineated above.
The operator $d^\dagger_k$ creates a doublon, while $h^\dagger_k$
creates a spin holon on the singly-occupied row. $t^d_{ij}$ and
$t^h_{ij}$ are the hopping matrices for the doublon and holon,
respectively: 
\beq\label{hopping-def}
t^d_{ij} &=& \alpha \tilde t^d_{ij} - (1-\alpha)  \Delta_{ij},\nonumber\\
t^{h\sigma}_{ij} &=& \alpha \tilde t^{h\sigma}_{ij} + (1-\alpha)
\Delta_{ij}, 
\eeq
where $\alpha = \left<{ (-1)^{\vec r_i} \vec S_i }\right>^{2m}$ for some positive $m$ is a function of the ordering parameter of the singly-occupied 
superlattice which is temperature dependent and $\tilde{J} = \alpha J$. In our calculation, $\alpha$ has the following form
\beq\label{alpha}
\alpha(T)=\left\{\begin{array}{ll}
0&,T>T_N\\
\left({\frac{T_N-T}{T_N} }\right)^m &,T<T_N
\end{array}\right.,
\eeq
and constitutes the only free parameter in our comparison with experiment.
Both $\alpha$ and the modified form of $J$ reflect the fact that the hopping processes depend on the antiferromagnetic ordering.  As a 
consequence, $J$ enters the energy cost only if antiferromagnetism is present. $ \Delta_{ij} = t ( \delta_{i_x \pm 1, j_x} \delta_{i_y,j_y} + 
\delta_{i_x, j_x} \delta_{i_y \pm 1, j_y})$ corresponds to the hopping mechanism for both the doublon and holon in the absence of 
antiferromagnetism. If the electron on the singly-occupied superlattice is antiferromagnetically correlated, the hopping mechanism for the doublon 
and holon becomes
\beq\label{hopping}
\tilde t^d_{ij} &=& - t \delta_{i_x \pm 1,j_x } \delta_{i_y \pm 1, j_y} ,\\
\tilde t^{h\sigma}_{ij} &=& + t( \delta_{i_x \pm 1, j_x }
\delta_{i_y, j_y} + \delta_{i_x, j_x} \delta_{i_y \pm 1, j_y}\nonumber\\& +&
\delta_{i_x \pm \sigma 1, j_x} \delta_{i_y \pm \sigma 1, j_y} ).
\nonumber \eeq
Likewise, $\alpha$ enters the diagonal hopping term because such
processes only enter if antiferromagnetism is present. Noting that
the creation of a doublon-holon pair on two adjacent rows costs a
total energy $2(V+J)$, we can write the non-interacting energies of
the doublon and holon as
\beq\label{free-energy} \varepsilon_{d,h} (\vec {k}) = V + \tilde J
+ t ^ {d, h} (\vec{k}), \eeq
with $t ^ d (\vec{k}) = - 2 \alpha t (\cos (k_x + \sqrt 3 k_y ) +
\cos (k_x - \sqrt 3 k_y )) - 2 t (1 - \alpha) (\cos(k_x) + \cos(
\sqrt 3 k_y))$ and $t ^ {h\sigma} (\vec{k}) = 2 t(\cos (k_x ) + \cos
(\sqrt 3 k_y )) + \alpha t\cos(k_x + \sigma \sqrt 3 k_y) $ are the
Fourier transforms of $t^d_{ij}$ and $t^{h\sigma}_{ij}$, respectively.  Localization 
corresponds to a binding of a spinon and a holon. The solution to
this model, obtained via a Bogoliubov rotation, has the
quasiparticle spectrum, \beq\label{qp-energy}
 E ^ - (\vec{k}) &=& + \cos ^2 \theta _k \varepsilon _d (\vec{k}) - \sin ^2 \theta _k \varepsilon _h (\vec{k}) -t\sin 2\theta_k\nonumber\\
 E ^ + (\vec{k}) &=& - \sin ^2 \theta _k \varepsilon _d (\vec{k}) + \cos ^2 \theta _k \varepsilon _h (\vec{k}) -t\sin 2\theta_k
\eeq
where $\tan 2\theta _k = -2t/(\varepsilon _d (k) + \varepsilon _h
(k))$ which corresponding to negative and positive charge excitation
respectively. As is clear from Fig. (\ref{band}), this model admits
a metal-insulator transition with a well-defined gap, $\Delta$,
between the localized and the extended states. This gap is zero for
$V+J-4t<0$ but otherwise given by \beq\label{delta} \Delta \approx
(V+J)(1-\frac{4t}{V+J}). \eeq
Experimentally, $J\approx 10meV$ and $t=0.2eV$\cite{hasan};
consequently, there is no solution to Eq. (\ref{delta}) if $V=0$ as
in the Hubbard model. However, $V$ is not known.  Hence, to
determine $\Delta$, we must rely on experiments. To describe the
experimental upturn of the resistivity at $53K$, $\Delta\approx
53K$, which results in $V\sim 880meV>4t$. This value of $V$ is
reasonable because even if we use the reduced value of the on-site
energy, $U/\sqrt{3}= 4.0eV/1.73=2.3eV$\cite{lee} reported recently
for the insulating state, we find that $V$ is of order $U/3$. The
importance of this estimate is that $V$ is large enough to stabilize
the charged ordered insulating phase.  We computed the resistivity
using the Kubo formula within the diagonalised quasiparticle spectrum in Eq. (\ref{qp-energy}).  Fig. (\ref{band}a) shows that the computed resistivity adequately reproduces the experimental trends. . The only free parameter is
$m$ in $\alpha$ which we set to 4. We have found, however, that the resistivity
is insensitive to $m$ as the divergence is governed primarily by the gap which is set by experiment. 
A key consequence of the insulating state proposed here is the
emergence of an antiferromagnetic superlattice with dimension
$a\times\sqrt{3}a$ on the singly occupied sites. In fact, this is
our key experimental prediction.  This can be established
rigorously. There is a weak antiferromagnetic coupling, $J_1$
between electrons separated by $\sqrt{3}a$ on neighbouring singly
occupied rows which is mediated by intervening doubly occupied
sites.  For this exchange interaction to be well-defined, doublons
should not be created; hence $V+J>4t$. Consequently, $J_1$ should be
a function of $V$.

To determine the rough magnitude of the effective spin coupling
constant, $J_1$, in the sublattice, we consider the $4^{th}$ order
perturbation process which flips the spins with respect to the charge
ordered state and obtain $J_1 = 4(t/V)^2J$. For the experiments at
hand, $V\approx .88eV\approx 4t$, implying that $J_1\approx J$.
Furthermore, the effective horizontal spin coupling constant, $J$,
should be renormalilzed by the same $4^{th}$ order process such that
$J\rightarrow J'=J+J_1\approx 1.25J$. Through the identical mechanism,
next-nearest neighbour sites (that is, sites along the diagonal) in
the singly occupied sublattice are also antiferromagnetically
coupled.  However, in this case, only {\it one} path exists.
Consequently, the next-nearest-neighbour exchange interaction is
$J_2=J_1/2$. Hence, the singly occupied sublattice obeys the
Heisenberg Hamiltonian
\beq H_{\rm singl. occup.}&=&J'\sum_{i}\textbf {S}_{i}\cdot
\textbf{S}_{i+\hat{x}}+J_1\sum_{i}\textbf{S}_{i}\textbf{S}_{i+\hat{y}}\nonumber\\
&+&J_2\sum_{\rm nnn}\textbf{S}_i\cdot \textbf{S}_j=H_0+J_2\sum_{\rm
nnn}\textbf{S}_i\cdot \textbf{S}_j \eeq
where $\hat{x} (\hat{y})$ denotes a nearest-neighbour along the
x-axis (y-axis). In the absence of $J_2$, the system develops 
long-range antiferromagnetic order for temperatures
less than $J_{\rm eff}\equiv\sqrt{JJ_1}\approx
13meV$\cite{mft}.  For $J_2\ne 0$,
antiferromagnetism obtains as long as $J_2<J_{\rm
eff}/2$\cite{singh} which is satisfied in our case.
 As the effective low-energy description of the
insulating state has now been reduced to a half-filled rectangular
lattice (by virtue of integrating out the doubly occupied
sublattice), particle-hole symmetry has now been reinstated.

Consequently, the Hall coefficient should vanish as $T\rightarrow 0$
for $n=1.5$ as is observed experimentally\cite{Princeton}. Once the
transition to the AF phase occurs, both the non-interacting
quasiparticles with negative and positive charge can contribute to
the Hall coefficient as mentioned previously. But, according to Fig.
\ref{band}, the negatively charged quasiparticles have a lower
energy. Hence, $R_H<0$ below the AF transition. Above $T_N$, no
order exists but the stripe is still stable. In this regime, the
sign of the Hall coefficient is determined strictly by the filling
which for $n=1.5$ should be hole-like. Since the Hall coefficient
vanishes at $T=0$, it is proportional to the charge difference
between particles and holes.  Hence, we computed the Hall
coefficient by integrating $2(f(E^+(k))-f(E^-(k))/e$ over all
momenta where $f$ is the Fermi function.  As Fig. (\ref{band}) indicates, the Hall coefficient contains has a sign change and a minimum at $T\approx 0.5J$ that are in superb quantitative agreement with experiment.  Likewise, the resistivity has a plateau where $R_H$ acquires its minimum value, as observed experimentally\cite{Princeton}). 
Here again we used $m=4$ and found that the overall qualitative features are insenstivive to the choice of $m$.  The robustness of the qualitative features of $R_H$ and the resistivity lend credence to Eq. (\ref{toy}) as the reductive model for the insulating state.

Within mean-field theory\cite{mft}, the upper bound to the
temperature at which there is a transition to this AF state is given by
$T_c=2J_{\rm eff}$. For reasonable system parameters, this energy
scale exceeds that of the gap for the insulating state and hence is
consistent with recent neutron scattering experiments\cite{lynn}
which find charge ordering and antiferromagnetism consistent with
Fig. (\ref{pstate}) above the temperature at which the resistivity
exhibits an upturn. The relative ordering of the energy scales is as
follows: the energy scale for destroying the stripe order is set by
the nearest-neighbour Coulomb energy $V$. Antiferromagnetic order is
destroyed above $J_{\rm eff}$. Between $\Delta$ and $J_{\rm eff}$,
AF order exists but the holons and doublons are mobile.
Antiferromagnetic order in the plane is maintained because diagonal
doublon hops do not generate ferromagnetic bonds but rather
exchanges doublons between the two sublattices. This ultimately
mixes the valence between the cob
alt atoms on the singly and doubly
occupied rows. Hence, rapid charge fluctuations on a time scale set by $\hbar/t\sim 10^{-14}s$ could be the origin
of the absence of two distinct Co environments reported in recent
NQR experiments\cite{mixing}. Nonetheless, such mixing ceases in the insulating
state as the holon and doublon are bound. Hence, our $T=0$ state 
contains two distinct Co environments.  High field NMR Co NMR experiments\cite{ning}
find, contrary to the NQR experiments, two distinct Co environments
corresponding with effective moments consistent with the charge ordered state
proposed here.

To summarize, in $Na_{0.5}CoO_{2}$, the ordering\cite{NeutronDiff,cohen} of the sodium ions
induces charge ordering in the $CoO_{2}$ layers so that both the
double occupancies and the singly occupancies reside on  different
tetragonal lattices.  However, charge localization obtains not from
pinning by the sodium superlattice but by  a delicate balance
between local Coulomb repulsions and the kinetic energy in the
CoO$_2$ plane.  The key prediction of this work that AF order on a
superlattice with dimension $a\times\sqrt{3}a$ has been confirmed
experimentally\cite{lynn}. Further, the diagonal hopping mechanism
identified here illustrates that the sign change of the Hall
coefficient at $T_N$ and the apparent itineracy\cite{mixing} of the
antiferromagnetism are intimately related.

\acknowledgments This work was supported by the NSF, Grant No.
DMR-0305864. We thank G. Baskaran for a useful e-mail exchange, T.
Vojta for a useful discussion on the Hall coefficient, E. Fradkin,
Patrick Lee and Steve Kivelson for correspondence which led to our
consideration of the impurity model in Eq. (\ref{toy}) and J. Lynn and Y. Lee for disclosing their experimental results.


\begin{thebibliography}{1}
\bibitem{j3}There is an additional ring process which does not involve the energy scale $U$ that can exchange the spins. Its amplitude is $O(t^3/V^2)$.
\end{thebibliography}
\end{document}